\documentclass[apj]{emulateapj}
\slugcomment{Accepted to ApJ 07/18/2010}
\usepackage{graphicx,amssymb,amsmath,times}
\usepackage{epsfig}
\usepackage{epstopdf}
\usepackage{rotating}

\def\cyg{Cygnus~X-3}

\begin{document}
\title{MAGIC constraints on $\gamma$-ray emission from Cygnus X-3}
\shorttitle{MAGIC constraints on $\gamma$-ray emission from Cygnus X-3}
\shortauthors{Aleksi\'c \emph{et al.}}

%
\author{
J.~Aleksi\'c$^{1}$,
L.~A.~Antonelli$^{2}$,
P.~Antoranz$^{3}$,
M.~Backes$^{4}$,
C.~Baixeras$^{5}$,
J.~A.~Barrio$^{6}$,
D.~Bastieri$^{7}$,
J.~Becerra Gonz\'alez$^{8,9}$,
W.~Bednarek$^{10}$,
A.~Berdyugin$^{11}$
% ...
(the MAGIC collaboration)\\
A.~Bulgarelli$^{29}$,
W.~Max-Moerbeck$^{30}$,
G.~Piano$^{31}$,
G.~Pooley$^{32}$,
A.~C.~S.~Readhead$^{30}$,
J.~L.~Richards$^{30}$,
S.~Sabatini$^{33}$,
E.~Striani$^{31}$,
M.~Tavani$^{31,33}$,
S.~Trushkin$^{34}$
}
\address{$^{1}$ IFAE, Edifici Cn., Campus UAB, E-08193 Bellaterra, Spain}
\address{$^{2}$ INAF National Institute for Astrophysics, I-00136 Rome, Italy}
\address{$^{3}$ Universit\`a  di Siena, and INFN Pisa, I-53100 Siena, Italy}
\address{$^{4}$ Technische Universit\"at Dortmund, D-44221 Dortmund, Germany}
\address{$^{5}$ Universitat Aut\`onoma de Barcelona, E-08193 Bellaterra, Spain}
\address{$^{6}$ Universidad Complutense, E-28040 Madrid, Spain}
\address{$^{7}$ Universit\`a di Padova and INFN, I-35131 Padova, Italy}
\address{$^{8}$ Instituto de Astrof\'{\i}sica de Canarias, E-38200 La Laguna, Tenerife, Spain}
\address{$^{9}$ Departamento de Astrof\'{\i}sica, Universidad, E-38206 La Laguna, Tenerife, Spain}
\address{$^{10}$ University of \L\'od\'z, PL-90236 Lodz, Poland}
\address{$^{11}$ Tuorla Observatory, University of Turku, FI-21500 Piikki\"o, Finland}
% ...
\address{$^{29}$ INAF-IASF, Bologna, Italy}
\address{$^{30}$ California Institute of Technology, Owens Valley Radio Observatory, Pasadena, CA 91125, USA}
\address{$^{31}$ Universit\`a Tor Vergata, and INFN, I-00133 Rome, Italy}
\address{$^{32}$ Cavendish Laboratory, Cambridge CB3 0HE, UK}
\address{$^{33}$ INAF-IASF, I-00133 Rome, Italy}
\address{$^{34}$ Special Astrophysical Observatory RAS, Nizhnij Arkhyz, Russia\\}
% ...

\begin{abstract}  
\cyg\ is a microquasar consisting of an accreting compact object orbiting 
around a Wolf--Rayet star. It has been detected at radio frequencies and 
up to high-energy $\gamma$ rays (above 100 MeV). However, many 
models also predict a very high energy (VHE) emission (above hundreds of GeV)
when the source displays relativistic persistent jets or transient ejections. Therefore, 
detecting such emission would improve the understanding of the jet physics. 
The imaging atmospheric Cherenkov telescope MAGIC observed \cyg\ for about
70 hr between 2006 March and 2009 August in different X-ray/radio spectral 
states and also during a period of enhanced $\gamma$-ray emission. MAGIC found 
no evidence for a VHE signal from the direction of the microquasar. An upper limit to 
the integral flux for energies higher than 250 GeV has been set to 
2.2~$\times$~$10^{-12}$~$\mathrm{photons}$~$\mathrm{cm^{-2}}$~$\mathrm{s^{-1}}$
(95$\%$ confidence level). This is the best limit so far to the VHE emission from this source.
The non-detection of a VHE signal during the period of activity 
in the high-energy band sheds light on the location of the possible 
VHE radiation favoring the emission from the innermost region 
of the jets, where absorption is significant. 
The current and future generations of Cherenkov telescopes may  
detect a signal under precise spectral conditions.

\end{abstract}


\section{Introduction}
\cyg\ is a bright and persistent X-ray binary, discovered in 
1966~\citep{Giacconi67}, but the high-energy processes 
occurring in the source are still poorly understood.  
It lies close to the Galactic plane at a distance between 
3.4 and 9.8~kpc, probably at 7~kpc, depending on different
distance estimates to the Cygnus OB2 association~\citep{Ling2009}. 
The nature and the mass of the compact object are still a
subject for debate. Published results suggest either a neutron 
star of 1.4~$M_{\odot}$~\citep{Stark2003} or a
black hole of less than 10~$M_{\odot}$~\citep{Hanson2000}. 
The identification of its donor star as a Wolf--Rayet 
star~\citep{Vankerkwijk1992} classifies it as a high-mass X-ray 
binary. Nevertheless \cyg\ shows a short orbital period of 4.8~hr, 
typical of the low-mass binaries, which has been inferred 
from the modulation of both the X-ray~\citep{Parsignault1972} and 
infrared emissions~\citep{Becklin1973}. 

Despite the strong X-ray absorption in this system, which 
may be caused by the wind of the companion star, the X-ray 
spectrum has been intensively studied. The source shows two 
main spectral X-ray states resembling the canonical states
of the black hole binaries: the hard state (HS) and the soft state 
(SS; Zdziarski $\&$ Gierlinski 2004; Hjalmarsdotter et al. 2008).  The HS is 
characterized by a weak soft thermal component and a strong 
non-thermal power-law emission peaking at hard X-ray energies,
whereas the SS, though showing a non-thermal tail~\citep{Szostek2008},
is dominated by the optically thick thermal disk emission. 
In Cygnus~X-3, however, the HS displays a high-energy 
cutoff at  $\approx$ 20 keV, significantly lower than the $\approx$ 
100 keV value found for black hole binaries~\citep[]{Hjalmarsdotter2004,
Zdziarski2004}. 

Adding to its peculiarity, \cyg\ is the brightest radio source among the 
X-ray binaries, quite frequently exhibiting huge radio flares~\citep[as 
intense as few thousand times the quiescent emission level of $\sim 
20$ mJy at 1.5 GHz;][]{Braes1972} first seen by \citet{Gregory1972}. During 
these outbursts, which occur mainly when the source is in the SS, 
and last several days, \cyg\ reveals the presence of collimated relativistic 
jets~\citep[e.g.,\ ][]{Marti2001,Mioduszewski2001,Geldzahler1983,
Miller-Jones2004}. Thus, \cyg\ has been 
classified as a microquasar. Based on arcsecond-scale radio 
observations and their relation with soft X-ray emission, six X-ray/radio
states have been identified: quiescent (flux densities $\sim$ 100 mJy), 
minor flaring ( $<$ 1Jy), suppressed ($<$ 100 mJy), quenched ($<$ 30 mJy), 
major flaring ($>$ 1 Jy), and post-flaring~\citep[]{Szostek2008,Tudose2009, Koljonen2010}. 
In the transition from the X-ray hard/radio quiescent state to the SS, 
the radio emission is strongly suppressed, and if it reaches the quenched 
level, the source usually produces a major radio flare~\citep{Waltman94}. 

\cyg\ has also historically drawn a great deal of attention due
to numerous claims of detection at TeV and PeV $\gamma$ rays, 
using early-days detectors in these energy ranges~\citep[]{Vladimirsky1973,
Danaher1981,Lamb1982,Dowthwaite1983,Samorski1983,Cawley1985,
Chadwick1985,Bhat1986}. However, a critical analysis of these 
observations raised doubts about their validity~\citep{Chardin1989}, and in 
recent years the more sensitive instruments have not confirmed those 
claims for energies above 500 GeV~\citep[]{Shilling2001,Albert2008a}.
Nevertheless, microquasars are believed to produce a very high energy 
(VHE) emission inside the jets: high density and magnetic fields provided by the 
accretion disk and by the companion star create favorable conditions 
for effective production of $\gamma$ rays~\citep[]
{Levinson1996,Romero2003,Bosch-Ramon2006}. This radiation 
could have either an episodic nature due to the ejection of strong
radio-emitting blobs~\citep{Atoyan1999}, generally occurring in the SS
in the case of \cyg\, or a (quasi)
stationary character being originated in the persistent compact jet present
during the HS~\citep[]{Bosch-Ramon2006}.   

Using data from the Energetic Gamma-Ray Experiment Telescope 
detector aboard the Compton Gamma-Ray Observatory,
\citet{Mori1997} reported an average flux of $(8.2 \pm 0.9) \times
10^{-7}$~photons cm$^{-2}$ s$^{-1}$ at energies above 100 MeV coming
from the direction of \cyg. However, no orbital modulation was detected in the signal,
precluding a solid association. The experimental situation in the high-energy 
region has been drastically changed by the results recently published by  
\emph{AGILE}~\citep{Tavani2009} and \emph{Fermi}/LAT~\citep{Abdo2009}. 
\emph{AGILE} detected the source in five different moments,
for a couple of days each, four of them corresponding to the peak
emissions shown in the detailed \emph{Fermi}/LAT light curve 
(see below). The last detection, occurred in 2009 July, has not 
been published yet by the \emph{AGILE} Collaboration (A. Bulgarelli et al., 
in preparation). On the other hand, \emph{Fermi}/LAT detected a 
clear signal from Cygnus X-3 above 100 MeV during two periods 
of enhanced activity lasting for several weeks and coinciding 
with the source being in the SS. The measured flux is variable 
and shows an orbital modulation, which confirms the origin 
of the signal from the microquasar. The \emph{AGILE} flux 
level is comparable with the one of the \emph{Fermi}/LAT 
flux peaks, as high as 2.0~$\times$~10$^{-6}$~photons~cm$^{-2}$ 
s$^{-1}$ above 100 MeV. 

Observations of binary systems with imaging atmospheric Cherenkov 
telescopes (IACTs) in the VHE band have proven very fruitful 
in recent years, with the detection of the orbitally modulated 
$\gamma$-ray emitters PSR\,B1259-63~\citep{Aharonian2005a}, 
LS\,5039~\citep{Aharonian2005b,Aharonian2006},
and LS~I~+61~303~\citep[]{Albert2006,Albert2008b,Albert2009,Acciari2008,
Acciari2009,Anderhub2009}. However, these systems may be different from 
Cygnus~X-3 since the radio and high-energy radiation could be produced 
in all three sources by the interaction of the winds of a star and an orbiting 
pulsar~\citep[]{Maraschi1981,Tavani1997,Dubus2006}. Nevertheless, \cyg\ 
GeV $\gamma$-ray modulation presents some common features to those
observed in LS\,5039 and LS~I~+61~303. On the other hand, 
although observations of other well-established microquasars, such as 
GRS~1915$+$105~\citep[]{Acero2009,Saito2009}, did not 
reveal any signal, there is an interesting possibility of VHE 
emission from Cygnus~X-1~\citep{Albert2007b}, still to be confirmed 
by an independent detection. The origin of this possible emission has been 
speculated to be linked to a maximum of the X-ray super-orbital 
modulation of the system~\citep{Rico2008}. Moreover, there is also 
a claim by \emph{AGILE} about a Cygnus~X-1 detection during a 
short flare (Sabatini et al. 2010, ATel 2512), which, however, has not been 
corroborated by \emph{Fermi}/LAT yet\footnote{Preliminary results can be found~at~\emph{
http://fermisky.vlogspot.com/2010/03/lat-limit-on-cyg-x-1-during-reported.html}}. 
Such experimental results endorsed by the theoretical predictions,
have encouraged deeper observations of the black hole binaries
in the VHE band.  

This paper reports observations of \cyg\ performed with the Major 
Atmospheric Gamma Imaging Cherenkov (MAGIC) telescope
between 2006 and 2009. All these observations were carried out
with the first stand-alone MAGIC telescope, MAGIC phase I.
\cyg\ observations were planned to cover different X-ray/radio states, 
including those that showed a strong high-energy $\gamma$-ray 
flux. This allowed us to search for VHE emission, above 250 GeV, 
from \cyg\ in the different cases for which this radiation is predicted in
theoretical scenarios. In addition, specific analyses were 
performed to look for predicted features, such as periodic and 
variable emission.
Section~\ref{sec::obs}  describes the performance of MAGIC
and the observational strategy. The analysis chain is explained 
in Section~\ref{sec::data}. The results obtained by MAGIC in a 
multi-wavelength context are reported in Section~\ref{sec::results}. 
A brief discussion is given in Section~\ref{sec::discussion}.
\section{Observations}\label{sec::obs}

MAGIC (phase I) is an IACT located at the Canary Island of La Palma 
(Spain), at 28$^\circ$.8 N--, 17$^\circ$.8 W, 2200 m above sea level. 
It is a 17 m diameter IACT with an energy threshold of 60 GeV 
(with the standard trigger). The collected Cherenkov light is 
focused on a multi-pixel camera composed of 576 
photomultiplier tubes (PMTs). 

The performance of the MAGIC telescope changed over the years. 
The largest improvement followed the 
upgrade of the 300 MHz readout system in 2007 February. The 
new multiplexed 2 GHz flash analog-digital converters improved 
the time resolution of the recorded shower images and reduced the 
contamination of the night sky background~\citep{Aliu2009}. 
Accordingly, the telescope integral flux sensitivity improved from 
$\approx 2.5\%$  to $\approx 1.6\%$ of the Crab Nebula flux for 
energies greater than 270 GeV in 50 hr of observations. At these 
energies, the angular and energy resolutions are $\approx 0^\circ$.1
and $\approx 20\%$, respectively. MAGIC can provide $\gamma$-ray 
point-like source localization in the sky with a precision of 
$\approx 2^\prime$~\citep{Albert2008c}. It is able to observe 
under moderate moonlight or twilight conditions~\citep[]{Albert2007a,
Britzger}, which, causing an increase of the night sky 
background, can be monitored through the PMT anode 
direct currents (DCs).

MAGIC pointed toward \cyg\ for a total of 69.2 hr between 
2006 March and 2009 August. Since the source is expected to 
be variable, the observations were  triggered by its state at 
other wavelengths. In 2006, observations were prompted by 
flares at radio frequencies. The MAGIC collaboration received two alerts from the 
RATAN-600 telescope on 2006 March 10 and on 2006 July 26 
(S. Trushkin, private communication 2006). Both radio flares had 
an X-ray counterpart: the source was in the SS according 
to \emph{RXTE}/ASM
and \emph{Swift}/BAT
measurements. 
In 2007, the observations were triggered using public 
\emph{RXTE}/ASM (1.5--12 keV) and \emph{Swift}/BAT 
(15--50 keV) data, as follows: (1) \emph{Swift}/BAT daily flux larger 
than 0.05 counts~$\mathrm{cm^{-2}}$~s$^{-1}$ and (2) ratio between \emph{RXTE}/ASM and 
\emph{Swift}/BAT count rates lower than 200. These criteria guaranteed
the source to be in the HS. During 2008 and 2009, 
MAGIC observed \cyg\ following high-energy $\gamma$-ray alerts issued
by the \emph{AGILE} satellite. The first of these alerts arrived in 2008 April 
18, the second one in 2009 July 18 (M. Tavani, private communication 2008; 2009). 
These last two campaigns occurred when the source was in the SS.
\begin{deluxetable}{ccccccc}[t]
\tablecaption{\cyg\ Observation log\tablenotemark{a}.
\label{tab::log}}
\tablehead{
\colhead{Obs.} & \multicolumn{2}{c}{Date} & \colhead{Eff. Time} & \colhead{Zd} & \colhead{DC\tablenotemark{b}} & \colhead{Spectral\tablenotemark{c}} \\
\colhead{Cycle} & (yyyy mm dd) & (MJD) &  (h)& $(^{\circ})$ & $(\mu$A) & \colhead{State}
}
\startdata
& 2006 03 23 & 53817 & 0.45 & 45--50&3.5 & \\
& 2006 03 24 & 53818 & 0.25 & 47--50&2.5 & \\
& 2006 03 26 & 53820 & 0.53 & 44--50&1.5 &\\
& 2006 03 28 & 53822 & 0.70 & 42--50&1.4 & \\
I & 2006 03 30 & 53824 & 0.80 & 41--50&1.3 & SS\\
& 2006 03 31 & 53825 & 0.90 & 40--50&1.4 & \\
& 2006 04 01 & 53826 & 1.00 & 38--50&1.8 & \\
& 2006 04 02 & 53827 & 0.92 & 40--50& 1.2& \\
& 2006 04 03 & 53828 & 1.05 & 38--50&1.2 & \\
\tableline	
& 2006 07 27 & 53943 &3.10 & 12--34&1.2 &\\
& 2006 07 28 & 53944 &2.53 & 12--31&1.1 & \\
II & 2006 07 29 & 53945 & 1.73 & 12--20&1.2 & SS\\
& 2006 07 30 & 53946 & 1.36 & 12--20&1.1& \\
& 2006 08 01 & 53948 & 0.92 & 12--20&1.4 &\\
& 2006 08 02 & 53949 & 1.88 & 12--19&1.2 & \\
\tableline
 & 2007 07 06 & 54286 & 2.16 & 19--45&2.3 & \\ 
& 2007 07 07 & 54287 & 4.53 & 12--45&2.3&\\
& 2007 07 08 & 54288 & 1.07  & 26--45&1.5 &\\
 & 2007 07 09 & 54289 & 0.38 & 34--40 &1.5 & \\
 & 2007 07 14 & 54295 & 1.82 & 12--21&1.5 & \\
 & 2007 07 15 & 54296 & 1.93 & 12--21&1.5 &\\
 & 2007 07 16 & 54297 & 1.93 & 12--21&1.4 & \\
 & 2007 07 24 & 54305 & 1.75& 12--23&1.4& \\
III& 2007 07 26 &  54307 & 0.98& 19--30&1.4 & HS\\
& 2007 07 27 & 54308 &  0.5 & 19--30& 1.4 & \\
& 2007 08 04 & 54316 & 0.73 & 27--35&1.5 & \\
& 2007 08 06 & 54318 & 2.72 & 12--30&1.6 &\\
& 2007 08 07 & 54319 & 1.93& 13--31&1.3& \\
& 2007 08 08 & 54320 & 1.58 & 14--31&1.4 & \\
& 2007 09 03 & 54346 & 1.86& 13--37& 2.5 & \\
\tableline
 & 2008 04 28 & 54584 &  0.31 & 32--40&2.9 & \\
IV& 2008 04 29 & 54585 &  1.07 & 24--41&3.2 &SS\\
& 2008 04 30 & 54586 &  1.33 & 23--40&2.6& \\
\tableline
 & 2009 07 19 & 55031 & 3.48 & 12--36 &1.2 &\\
& 2009 07 21 & 55033 &  2.21 & 14--42&1.2& \\
V & 2009 07 22 & 55034 & 1.63 & 12--27&1.1 & SS\\
& 2009 08 01 & 55044 & 1.81 & 17--39&1.5& \\
& 2009 08 02 & 55045 & 0.88 & 27--41&1.6 & 
\enddata
\tablenotetext{a}{From left to right: observational cycle, date of the 
beginning of the observations, also in MJD, effective time after quality cuts,
zenith angle range, the anode PMT DCs, and the spectral state.\\}
\tablenotetext{b}{The anode PMT DCs show the level
of the moonlight conditions.\\}
\tablenotetext{c}{Spectral state was derived by using \emph{Swift}/BAT and 
\emph{RXTE}/ASM data.}
\end{deluxetable}%

At La Palma, \cyg\ culminates at a zenith angle of $12^\circ$. The 
observations were carried out at zenith angles between 12$^\circ$ and 
50$^\circ$. They were performed 
partially in a on--off mode and in a false-source tracking mode 
(wobble)~\citep{Fomin1994}, the latter pointing to two directions at 
$24^\prime$ distance and opposite sides of the source.
The entire \cyg\ data set recorded by MAGIC amounts to 
69.2 hr, out of which 12.5 hr\footnote{9.6 hr were rejected 
due to high-altitude Saharan dust (calima), which affects the 
atmosphere above Canary islands and is more intense during 
summer.} were rejected from further analysis because of their 
anomalous event rates, leading to a total of 56.7 hr of useful 
data distributed in 39 nights of observation. The details of the 
observations are quoted in Table~\ref{tab::log}.\\

\section{Data Analysis}\label{sec::data}
Data analysis was carried out using the standard MAGIC calibration 
and analysis software: once the PMT signal pulses are calibrated 
\citep{Albert2008d}, pixels containing no useful information for 
the shower image reconstruction are discarded. This is done by an 
image cleaning procedure which takes into account the amplitude 
and the timing information of the calibrated signals~\citep{Aliu2009}. 
The constraints on the signal amplitude are increased 
in the case of moonlight conditions, when the pixel DCs become larger 
than 2.5 $\mu$A (see Table~\ref{tab::log}). This higher image cleaning 
allows us to use the standard Monte Carlo (MC) during the analysis of 
these data, without the need of any correction, as long as the average 
pixel DCs are below 4 $\mu$A~\citep{Britzger}.

Afterward, the surviving pixels are used to compute the Hillas event 
image parameters~\citep{Hillas1985}. In addition, Hillas and timing 
variables are combined into a single $\gamma$/hadron discriminator
(\emph{hadronness}) and an energy estimator by means of the Random 
Forest classification and regression algorithm, which takes into account the correlation 
between the different Hillas and timing variables~\citep{Breiman2001,
Albert2008e}. These algorithms are trained with a sample of 
MC-simulated $\gamma$-ray events and a sample of background events 
extracted from real data. The used timing variables allow an improvement 
of the analysis sensitivity of a factor of 1.4 (1.2) in data recorded with the new 
ultra-fast (old 300 MHz) readout system~\citep{Aliu2009}. 

Images with a total charge below 200 photo-electrons were discarded from 
further analysis in order to homogenize the different data and MC samples
and achieve a common stable energy threshold of all the analyses. 
The number of $\gamma$-ray candidates in the direction 
of the source is estimated by using the distribution of \emph{alpha} 
angle~\citep{Hillas1985}, which is related to the shower orientation. 
The evaluation of the background depends on the data taking mode. 
For data taken in the on--off mode in which the signal data sample is 
called ``on" data, the background is estimated using a different sample, 
called ``off" data. The latter has a similar amount of data recorded during 
the same epoch as the ``on" sample at similar zenith angles and 
atmospheric conditions with no known $\gamma$-ray source in the 
field of view. For the observations carried out in the wobble
mode, the background is extracted from three circular control regions, 
located symmetrically to the source position with respect to the camera 
center. 

The \cyg\ data set extends over five different, one-year long cycles 
of observation, which are characterized by varying performances of the 
telescope. In addition, cycle I data were recorded in the on--off mode, all the 
others in the wobble mode, cycle IV data were taken under moderate 
moonlight conditions. 

Each cycle of data was analyzed separately using the appropriate image
cleaning procedure, and a matching sample of MC-simulated $\gamma$-ray 
events. The analysis selection cuts, on \emph{hadronness} and \emph{alpha}, 
for each cycle of data were obtained optimizing the sensitivity in a Crab
Nebula sample and requiring at least 70$\%$ $\gamma$-ray selection efficiency.
The selected Crab Nebula sample was recorded during the same cycle 
at similar zenith angles and in the same data taking mode. 

All the analyses were then combined in order to calculate the integral 
flux upper limits (ULs) for energies greater than 250 GeV, which is the 
energy threshold of each cycle of data. 
The obtained ULs at the 95$\%$ confidence level (CL) were computed 
after \citet{Rolke} method assuming a Gaussian 
background and 30$\%$ of systematic uncertainties in the flux 
level~\citep{Albert2008c}. The spectrum was assumed to be 
Crab like~\citep{Aharonian2004}, with a photon index of~2.6. 
However, a 30$\%$ change in the photon index yields a 
variation of less than 1$\%$ in the ULs.

\section{Results}\label{sec::results}
\begin{figure}[!b]
\begin{center}
\epsscale{1.25}
\plotone {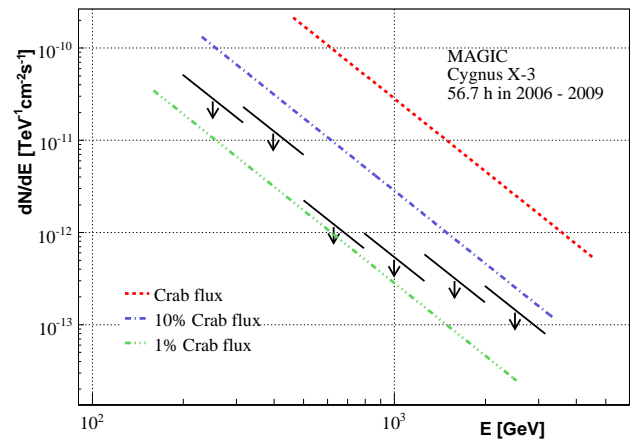} 
\figcaption{Differential flux ULs at 95$\%$ CL for the VHE 
time-integrated emission. The slope of the arrows indicates the 
assumed power-law spectrum with a photon index of 2.6.
\label{fig::DiffULtot}}
\end{center}
\end{figure}
\begin{figure*}[!t]
\begin{center}
\epsscale{1.2}
\plotone{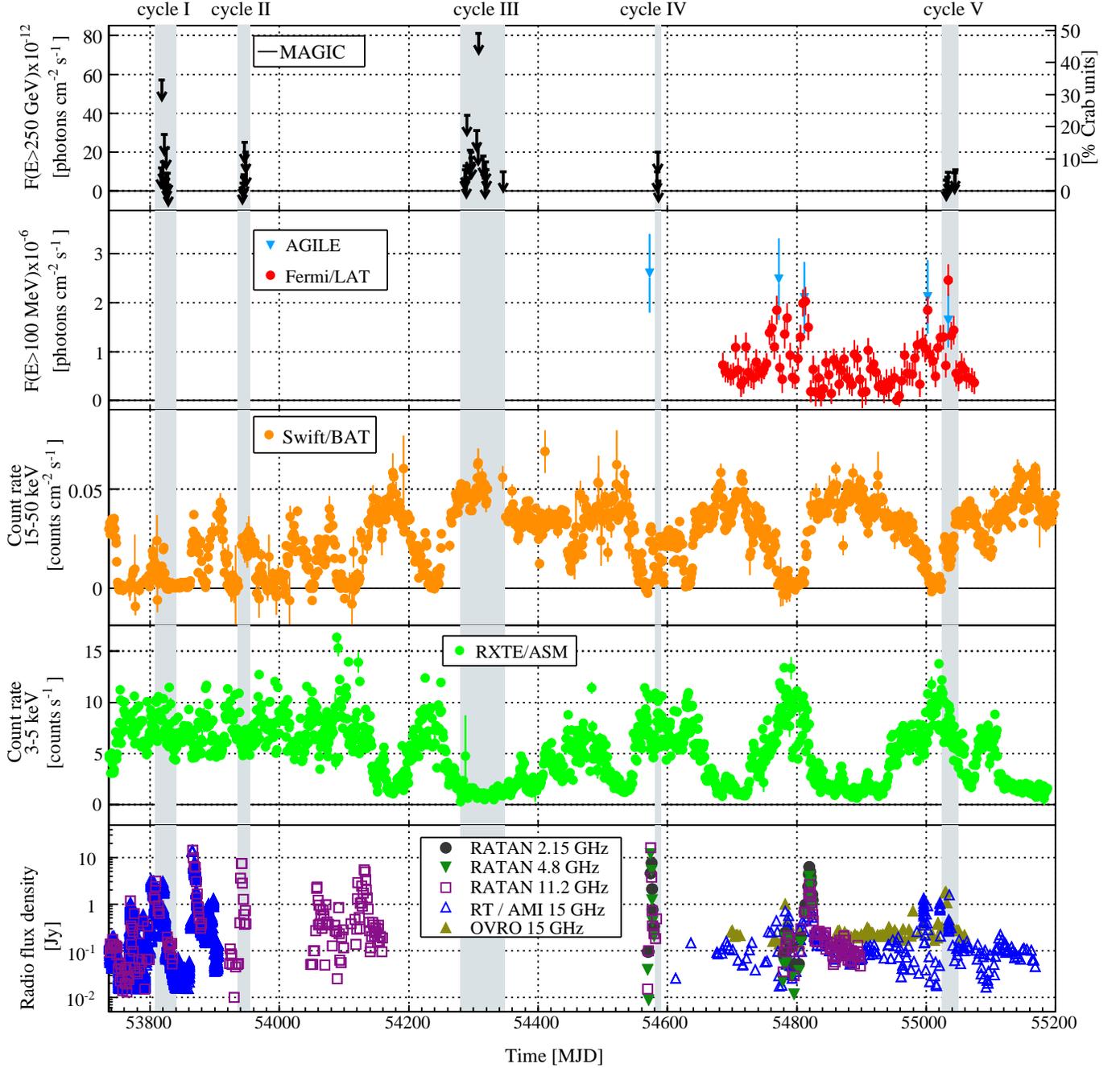} 
\figcaption{From top to bottom: daily MAGIC VHE integral flux ULs for 
$\mathrm{E}$ $>$ 250 GeV, high-energy $\gamma$-ray (\emph{AGILE} 
and \emph{Fermi}/LAT), hard X-ray (\emph{Swift}/BAT), soft X-ray 
(\emph{RXTE}/ASM), and radio fluxes measured for \cyg\ as a function of 
time (from 2006 January 1 until 2009 December 15). The grey bands show 
the periods corresponding to the MAGIC observations. The OVRO and AMI 
15 GHz data generally agree well, except for the offset apparent during 
steady periods which is due to unrelated extended emission resolved out by AMI.  
\label{fig::multi}}
\end{center}
\end{figure*}

A search for VHE $\gamma$-ray emission from \cyg\ was performed separately 
for each cycle of observations. None of them yielded a significant excess. 
The combination of all the data samples yields a 95$\%$ CL
upper limit to an integral flux of 
2.2~$\times$~$10^{-12}$~photons~cm$^{-2}$~s$^{-1}$ 
for energies greater than 250 GeV. It corresponds to 1.3$\%$ of the Crab 
Nebula flux at these energies. The differential flux ULs are shown in 
Table~\ref{tab::DiffULtot} and in Figure~\ref{fig::DiffULtot}. 

Given that \cyg\ flux is variable at other energy bands on time scales of days, 
$\gamma$-ray signals were searched for also on a daily basis. 
No significant excess events were found in any observation night.
The integral flux ULs for energies above 250 GeV are shown 
in Table~\ref{tab::ULdaily} and in the top panel of Figure~\ref{fig::multi}. 
\begin{figure*}[!t]
\begin{center}
\epsscale{1.2}
\plotone{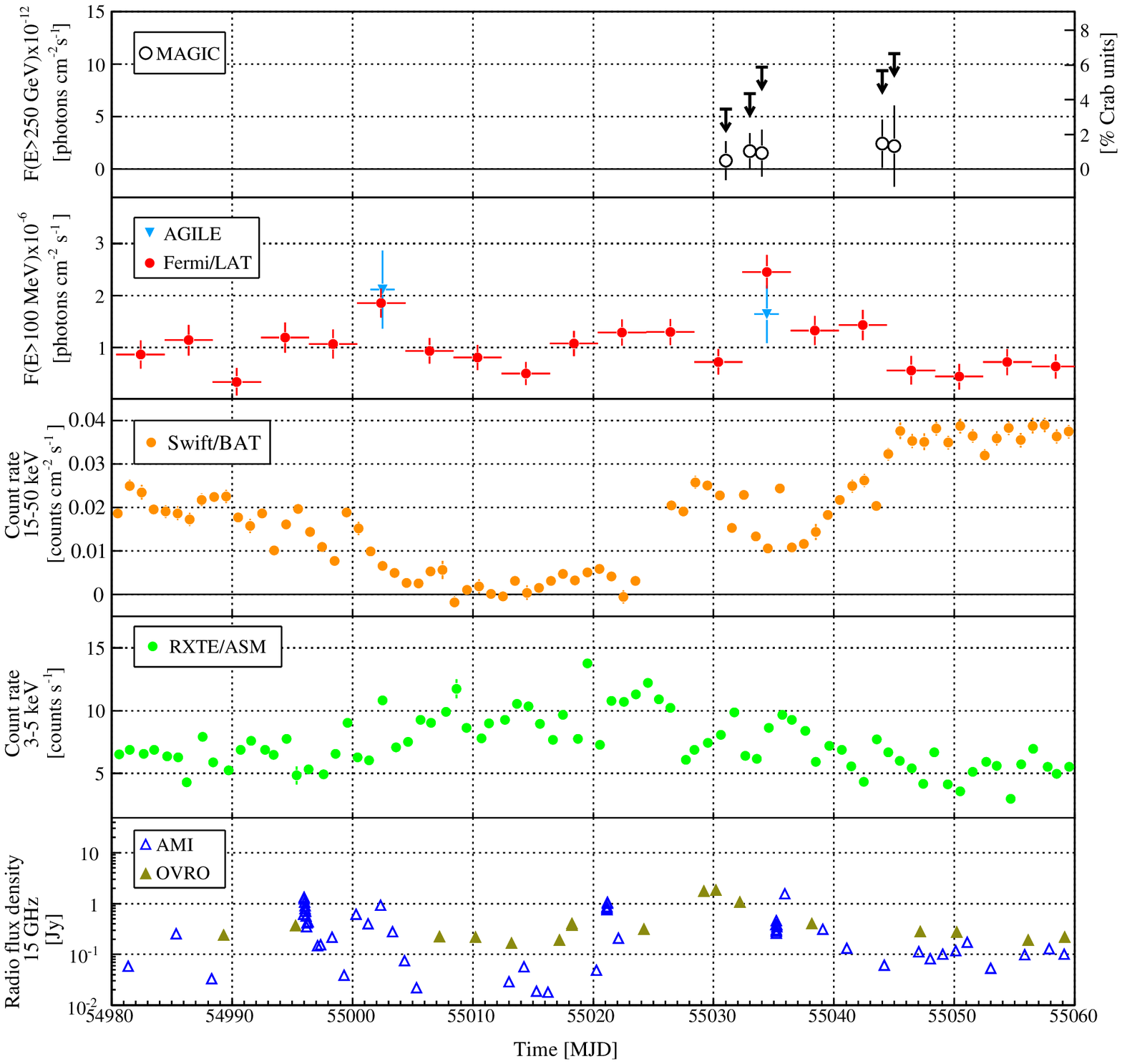} 
\figcaption{Zoom of Figure~\ref{fig::multi} around the cycle~V campaign 
between 2009 May 29 and August 17. 
The open black points in the VHE MAGIC panel show
the non-significant measured integral fluxes with their statistical error bars 
(whereas the ULs take into account also the systematic errors).
\label{fig::zoom5}}
\end{center}
\end{figure*}

In Figure~\ref{fig::multi}, MAGIC results are presented in a multi-wavelength 
context where the above-mentioned flux variability at different energy bands 
is rather clear. In particular, it displays daily MAGIC VHE integral
ULs, high-energy $\gamma$-ray (\emph{AGILE} and \emph{Fermi}/LAT 
(0.1--30 GeV)), hard X-ray (\emph{Swift}/BAT (15--50 keV)), soft X-ray 
(\emph{RXTE}/ASM (3--5 keV)), and radio measured fluxes from 2006 
January 1 (MJD 53736) until 2009 December 15 (MJD 55180). 
The soft X-ray energy band of \emph{RXTE}/ASM is between 1.5 and 12 keV,
out of which only the 3--5 keV band was used, as it yields to the cleanest 
radio/X-ray correlation~\citep{Szostek2008}. The radio measurements, displayed 
in logarithmic scale, were provided by the RATAN-600 telescope at 2.15, 4.8, and 
11.2 GHz and by the Ryle telescope (RT), the Arcminute Microkelvin Imager 
(AMI) telescope and the Owens Valley Radio Observatory (OVRO) 
40 meter telescope at 15 GHz. 

The soft and hard X-ray fluxes shown in Figure~\ref{fig::multi} allow us to derive the 
X-ray spectral state of \cyg\ during MAGIC observations. Cycle III data
are the only ones taken when the source was in the HS, as requested
by the observational trigger. All the other observations were carried out when 
the source was in the SS. The first two MAGIC observational
campaigns (cycles I and II) were triggered by flares at radio frequencies, 
which are expected when the source is in the SS. Unfortunately, the
conditions on the trigger together with MAGIC constraints on observation time did 
not allow a simultaneous coverage of the flaring states: MAGIC started pointing
toward \cyg\ twelve and two days after the strong radio emission, respectively. 
On the other hand, in the last observational cycles (IV and V), MAGIC observed 
the source during the SS, following two \emph{AGILE} alerts on a 
high-flux activity in the high-energy band: the GeV emission occurs only in this 
X-ray spectral state. During the second alert, in 2009 July (MJD 55030), 
MAGIC recorded some data simultaneous with a GeV peak emission and  
did not detect any VHE signal. This important result will be discussed in detail 
in Section~\ref{sec::high-energy}.

Microquasars are expected to produce VHE emission inside the jets either
when they are compact and persistent, mainly in the HS, or in the
presence of strong radio-emitting-blobs, which most likely happen during 
an SS. Therefore, the results of the MAGIC observations were divided
according to the X-ray spectral state of the source, as described in 
Sections~\ref{sec::SS} and \ref{sec::HS} for the SS and HS, 
respectively. Besides, in the scenario where the VHE radiation is emitted
during powerful synchrotron radio ejections, there might be a correlation 
between these two wavelengths~\citep{Atoyan1999}. Therefore, in 
Section~\ref{sec::X-ray/radio} the X-ray/radio states during the MAGIC 
observations are quoted.

\subsection{Results during High-energy $\gamma$-ray Emission}
\label{sec::high-energy}
\citet{Abdo2009} presented the \emph{Fermi}/LAT observations 
of \cyg\ between 2008 August 4 and 2009 September 2. 
They detected a strong signal above 100 MeV, with an overall significance 
of more than 29 standard deviations, which is mainly dominated by 
two active periods: MJD 54750--54820 and 54990--55045. These 
active periods coincide with the X-ray SS of the source, indicating 
that \cyg\ emits high-energy $\gamma$-rays during this spectral state, 
with an average flux of 
1.2~$\times$~$10^{-6}$~photons~cm$^{-2}$~s$^{-1}$
and a photon index of 2.70~$\pm$~0.05$_{\mathrm{stat}}$
$\pm$~0.20$_{\mathrm{sys}}$ (under the assumption of a 
power-law spectrum). They also estimated that the peak flux can be as 
high as 2~$\times$~$10^{-6}$~photons~cm$^{-2}$~s$^{-1}$
without providing any photon index. 

The five \emph{AGILE} detections between 100 MeV and 3 GeV (Tavani et al. 2009; 
A. Bulgarelli et al., in preparation) coincide with the strongest high-energy
outbursts, which can be seen overlapping them with the \emph{Fermi}/LAT 
light curve. The \emph{AGILE} integral flux averaged over the first four detections 
is estimated to be 1.9~$\times$~$10^{-6}$~photons~cm$^{-2}$~s$^{-1}$ 
with a corresponding photon index of 1.8~$\pm$~0.2.

In cycle~V, MAGIC pointed at \cyg\ during the second period 
of high-energy enhanced activity detected by \emph{Fermi}/LAT, as shown 
in Figure~\ref{fig::zoom5}.   
In particular, MAGIC carried out observations  simultaneous with a 
GeV emission peak on 2009 July 21 and 22 (MJD 55033--55034), 
and it did not detect any VHE emission. The corresponding MAGIC 
integral flux ULs above 250 GeV are lower than $6\%$ of the Crab 
Nebula flux. 

\begin{figure}[!h]
\begin{center}
\epsscale{1.25}
\plotone {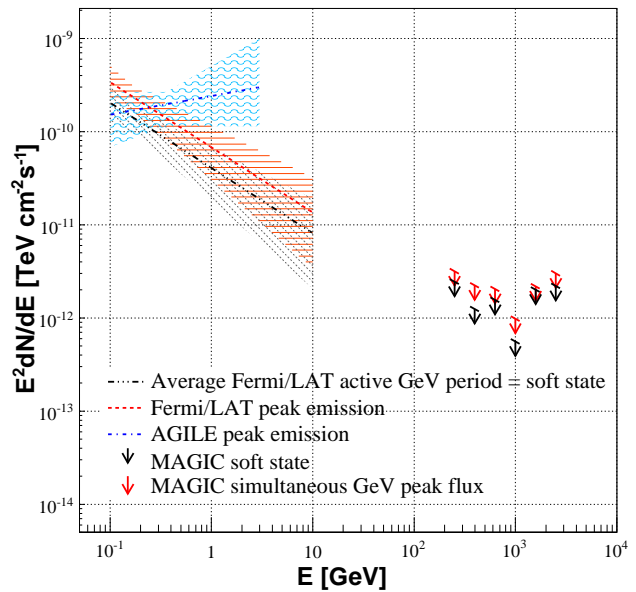} 
\figcaption{\cyg\ SED in the high-energy and VHE bands.
The lines indicate the power-law spectra derived from \emph{Fermi}/LAT 
and \emph{AGILE} integral fluxes and photon indices, where the corresponding
errors were taken into account and are shown in shadowed areas.
The arrows display the 95$\%$ CL MAGIC differential flux ULs and their slope 
indicates the assumed power-law spectrum with a photon index of 2.6.
The black indicators show the SED during the period of enhanced GeV activity 
coinciding with the SS, whereas, the red and blue ones during 
the high-energy peak emission (MJD 55031--55034).  
\label{fig::fermi}}
\end{center}
\end{figure}

Figure~\ref{fig::fermi} shows the spectral energy distribution (SED) of \cyg\
between 100 MeV and 5 TeV including MAGIC 95$\%$ CL upper limits 
at VHE, and the power-law spectrum in the high-energy range 
reported by both \emph{AGILE} and \emph{Fermi}/LAT. The 
spectra take into account the error on the photon index and the one on the
integral flux, which is 30$\%$ and 40$\%$ for \emph{Fermi}/LAT and \emph{AGILE},
respectively. The SED for the average SS was obtained 
considering the average \emph{Fermi}/LAT flux and MAGIC results of
the SS data set. On the other hand, to obtain the SED during a high-energy emission 
peak, the MAGIC measurements simultaneous with the GeV emission peak 
and both \emph{Fermi}/LAT and \emph{AGILE} spectral power-law fits were used. 
The \emph{Fermi}/LAT photon index during the peak was assumed to be 
the same as the one for the SS average flux. Being the latter dominated 
by the brightest phases of the $\gamma$-ray outburst, it can also be considered 
representative of the flaring activity. 
Both MAGIC ULs are consistent with the extrapolation of the 
\emph{Fermi}/LAT spectra up to VHE, but not with the extrapolation of the 
harder \emph{AGILE} spectrum, which would suggest a cutoff between
some tens and 250 GeV. 

\subsection{Results During the Soft State}
\label{sec::SS}

The MAGIC telescope pointed at \cyg\ when it was in the SS
during the observational cycles I, II, IV and V, corresponding 
to a total of 30.8 hr. For these observations, soft X-ray 
measurements in the 3--5 keV band are always above the transitional 
level set at 3 counts s$^{-1}$ by \citet{Szostek2008}. 

\begin{figure*}[!t]
\begin{center}
\epsscale{0.58}
\plotone{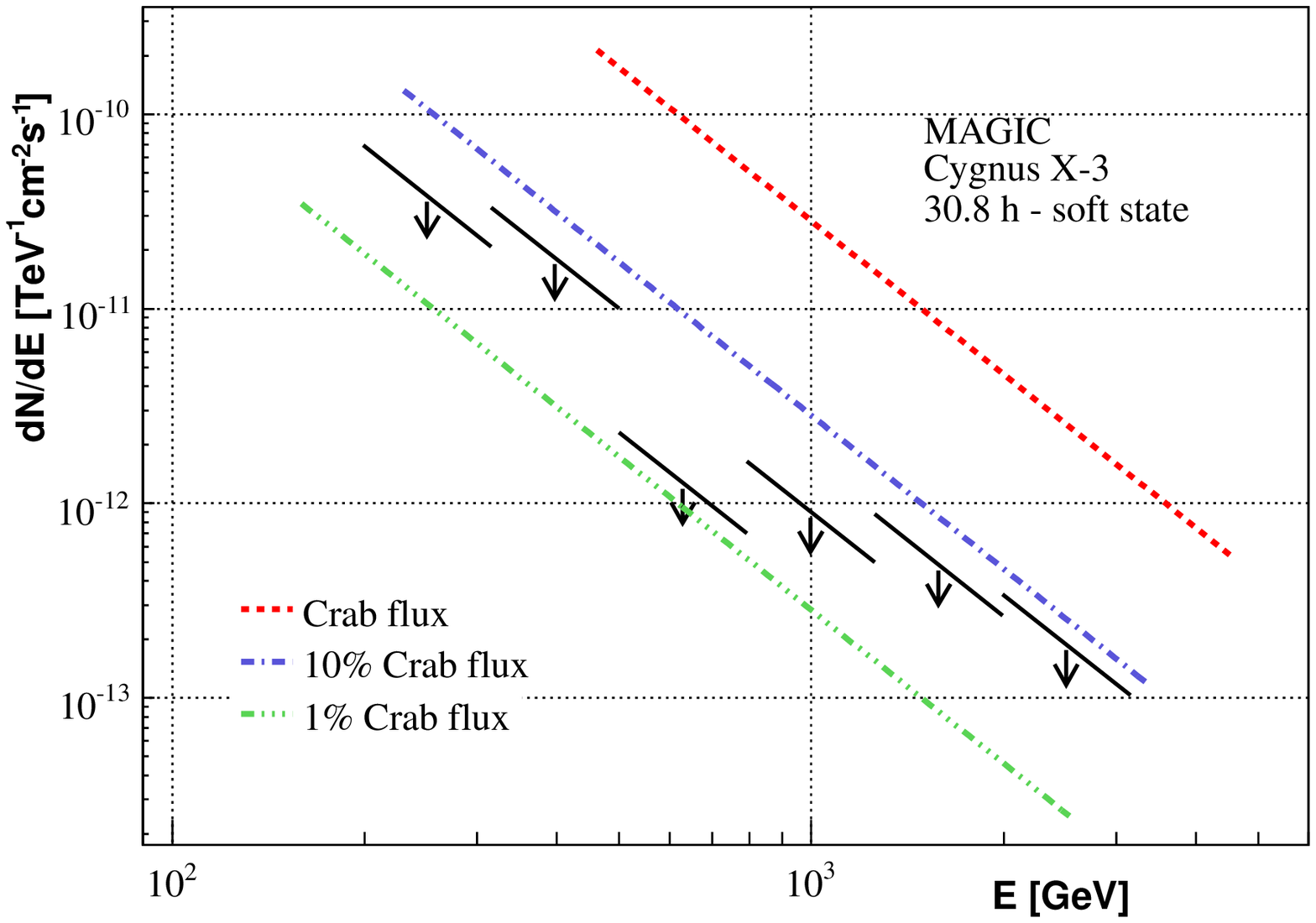} 
\plotone{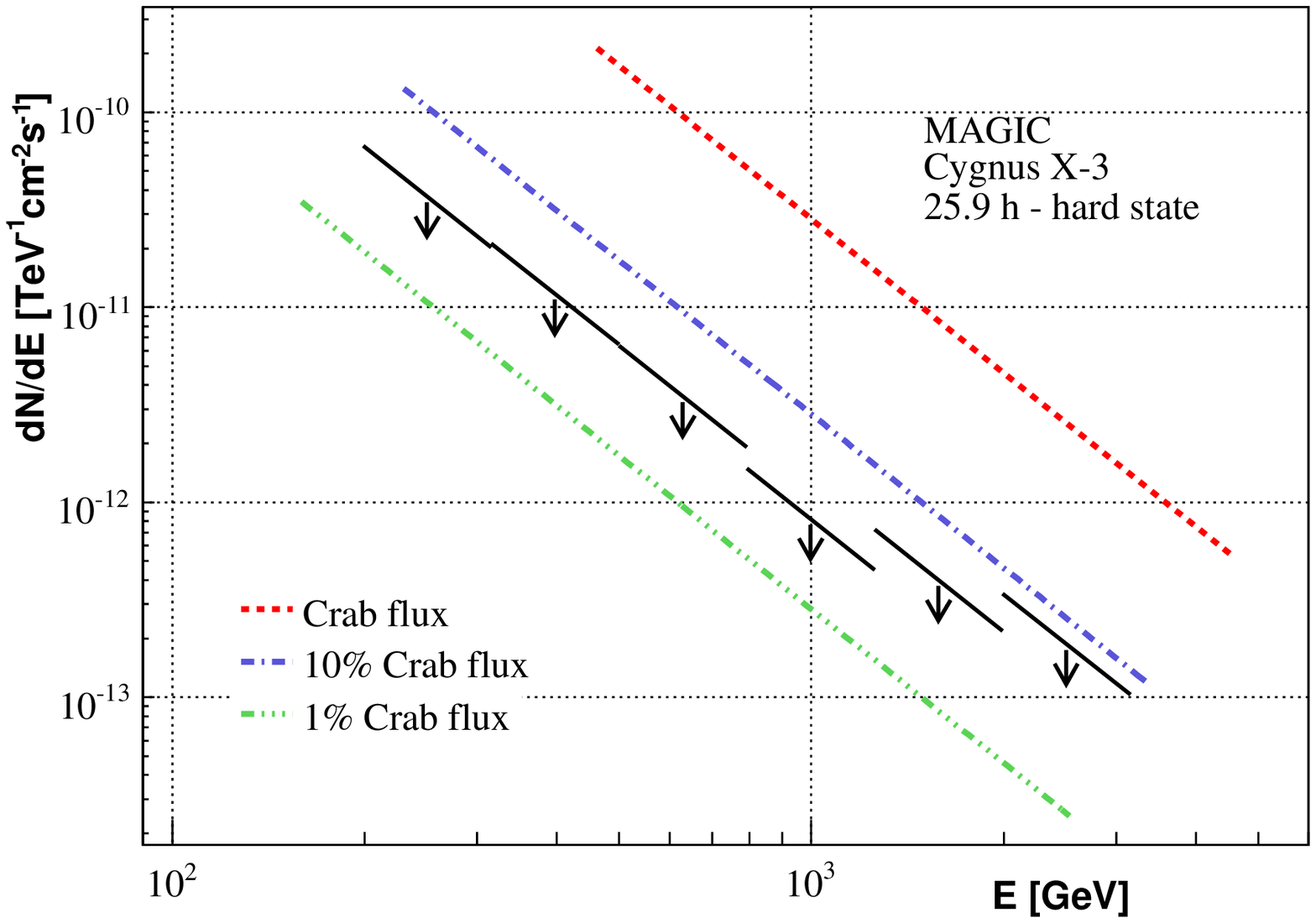} 
\figcaption{Ninety-five percent differential flux ULs for the SS (left panel) 
and HS (right panel) observations. The slope of the arrows 
indicates the assumed power-law spectrum with a photon index of 2.6.
\label{fig::DiffULspectral}}
\end{center}
\end{figure*}

After having analyzed each data cycle separately, the data sets 
were combined in order to obtain a global UL to the integral flux 
for the SS. The UL at 95$\%$ CL for energies 
greater than 250 GeV is 
4.1~$\times$~$10^{-12}$~photons~cm$^{-2}$~s$^{-1}$,
i.e., $\sim$~2.5$\%$ of the Crab Nebula flux. The differential flux 
ULs for this spectral state are shown in Table~\ref{tab::DiffULspectral}
and in the left panel of Figure~\ref{fig::DiffULspectral}.
\label{sec::HS}
\begin{figure*}[!t]
\epsscale{0.58}
\begin{center}
\plotone {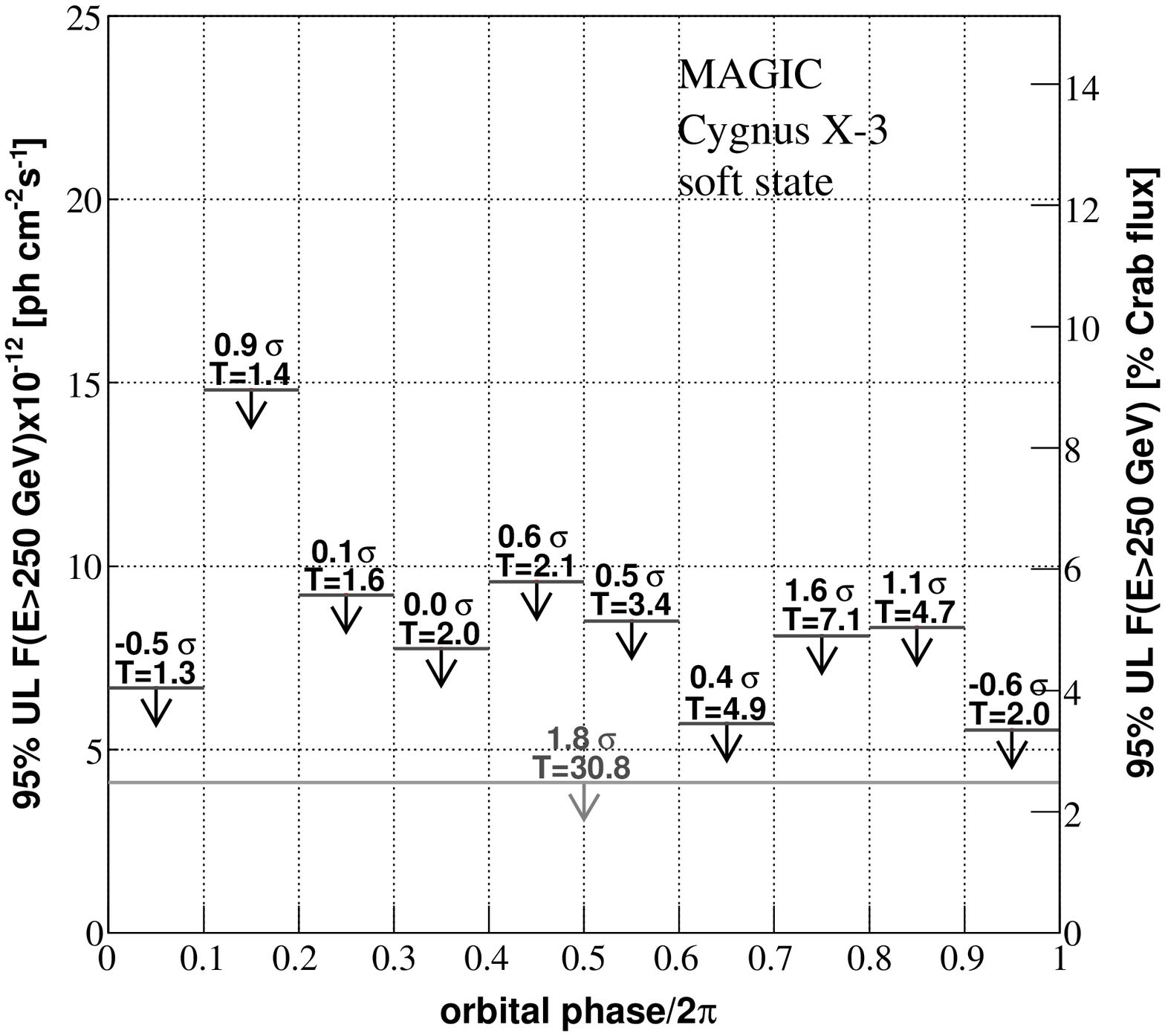} 
\plotone {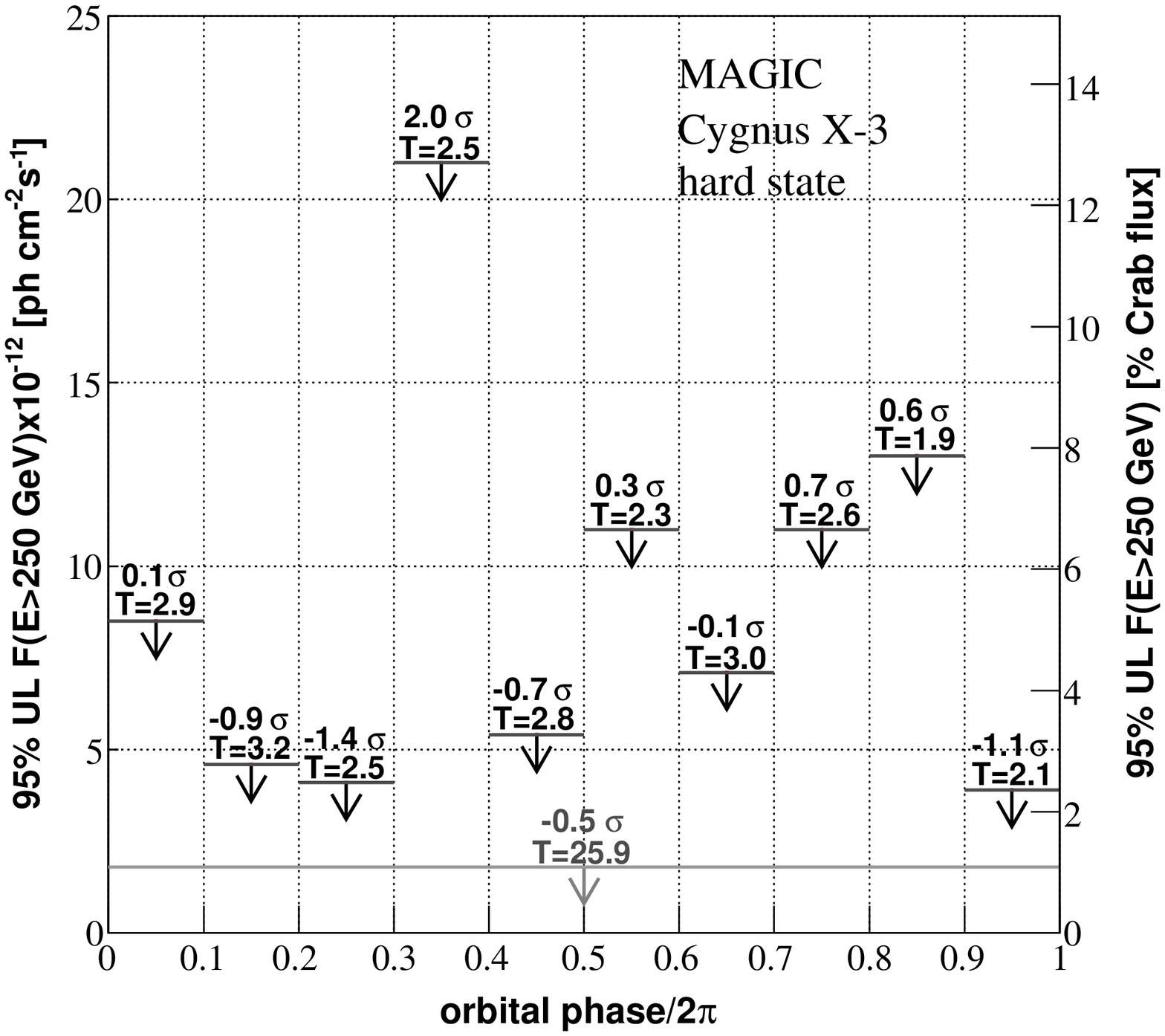} 
\figcaption{Phase-wise integral flux ULs for $E$ $>$ 250 GeV for the SS (left panel)
and HS (right panel). The effective observation time (in hours) and the signal 
significance for each phase bin are written on top of each arrow. The gray arrow indicates 
the integral flux UL on the VHE time-integrated emission.\label{fig::PhaseULspectral}}
\end{center}
\end{figure*}

Due to the highly anisotropic radiation from the companion star, the 
predicted $\gamma$-ray emission above 250 GeV would be modulated according to 
the orbital phase~\citep{Bednarek1997}. Absorption might play
an important role in making the VHE orbital modulation difficult to be detected 
by the current sensitivity instrumentation~\citep{Bednarek2010}. 
The orbital modulation of the GeV $\gamma$-ray emission was detected 
only when the source was in the SS~\citep{Abdo2009}. MAGIC searched 
for such modulation in this spectral state. A phase-folded analysis was performed 
assuming the parabolic ephemeris in~\citet{Singh2002}. The results are 
shown in the left panel of Figure~\ref{fig::PhaseULspectral} and in 
Table~\ref{tab::PhaseULspectral}.  No evidence of VHE $\gamma$-ray 
signal was found in any phase bin. The obtained integral flux ULs are 
smaller than 10$\%$ of the Crab Nebula flux for all of them.

\subsection{Results During the Hard State}
\label{sec::HS}

The 25.9 hr of MAGIC cycle III data sample were obtained 
when \cyg\ was in the HS. \emph{Swift}/BAT count rates 
during this cycle are rather high, greater than 0.05 
counts~cm$^{-2}$~s$^{-1}$, whereas the soft X-ray fluxes 
in the 3--5 keV band are below 2 counts~s$^{-1}$.  

For this spectral state, the integral flux UL for energies greater than 
250 GeV is 1.1$\%$ of the Crab Nebula flux (1.8~$\times$~$10^{-12}$~photons
~cm$^{-2}$~s$^{-1}$). 
The differential flux ULs are quoted in Table~\ref{tab::DiffULspectral}
and plotted in the right panel of Figure~\ref{fig::DiffULspectral}.
The performed phase-folded analysis for this spectral state did not yield 
any significant detection. The computed ULs to the integral flux are, 
on average, at the level of 5$\%$ of the Crab Nebula flux (see the right panel of 
Figure~\ref{fig::PhaseULspectral} and Table~\ref{tab::PhaseULspectral}).

\subsection{Results During X-ray/Radio Rtates}
\label{sec::X-ray/radio}

\begin{figure}[!h]
\begin{center}
\epsscale{1.25}
\plotone {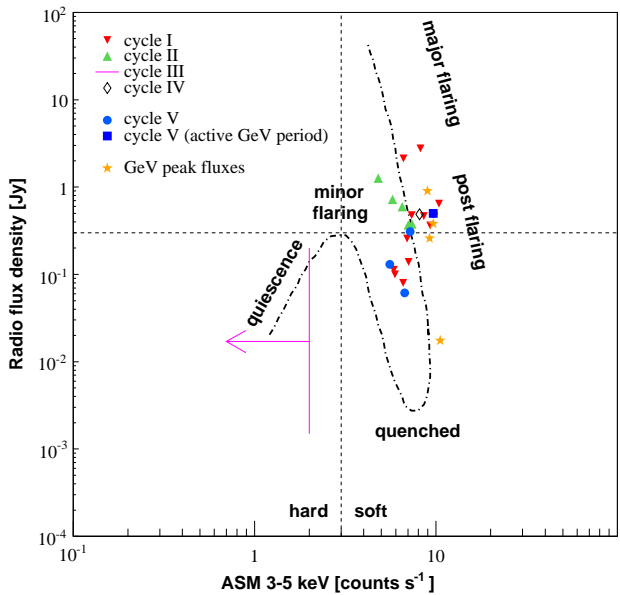} 
\figcaption{Soft X-ray counts vs. radio fluxes simultaneous with the
GeV peak fluxes (yellow stars) and MAGIC observations, where the different marker
colors identify the five MAGIC observational cycles. The pink arrow shows the X-ray
flux level during the cycle III MAGIC campaign, for which radio data are not available.
Radio measurements are at 11.2 GHz from RATAN-600 and at 15 GHz from RT/AMI. 
The dot-dashed line shows the expected ``saxophone" shape, adapted from~\citet{Szostek2008}. 
\label{fig::saxo}}
\end{center}
\end{figure}

MAGIC observed \cyg\ in both X-ray main spectral states (see 
Sections~\ref{sec::SS} and \ref{sec::HS}). However, the state of 
the source can be further characterized by simultaneous radio flux. 
\citet{Szostek2008} identified six different X-ray/radio states studying 
simultaneous observations of the Green Bank Interferometer at 8.3 GHz and 
\emph{RXTE}/ASM in the energy range 3--5 keV. The relation between 
these two energy bands is shown in the so-called saxophone plot (see 
Figure~\ref{fig::saxo}). It was noted that the use of other radio frequencies 
yields similar results. This gives us the confidence that a direct comparison
between their and our results using 15 GHz (RT and AMI) and 
11.2 GHz (RATAN-600) is reasonable. The OVRO and AMI 15 GHz data 
generally agree well, but for an $\sim$0.12 Jy offset apparent during steady 
periods, probably due to unrelated extended emission resolved out by AMI. 
Thus, only the AMI 15 GHz were used in this analysis, although our 
conclusions are not substantially affected by this choice.

\begin{figure*}[h]
\epsscale{1.2}
\begin{center}
\plotone {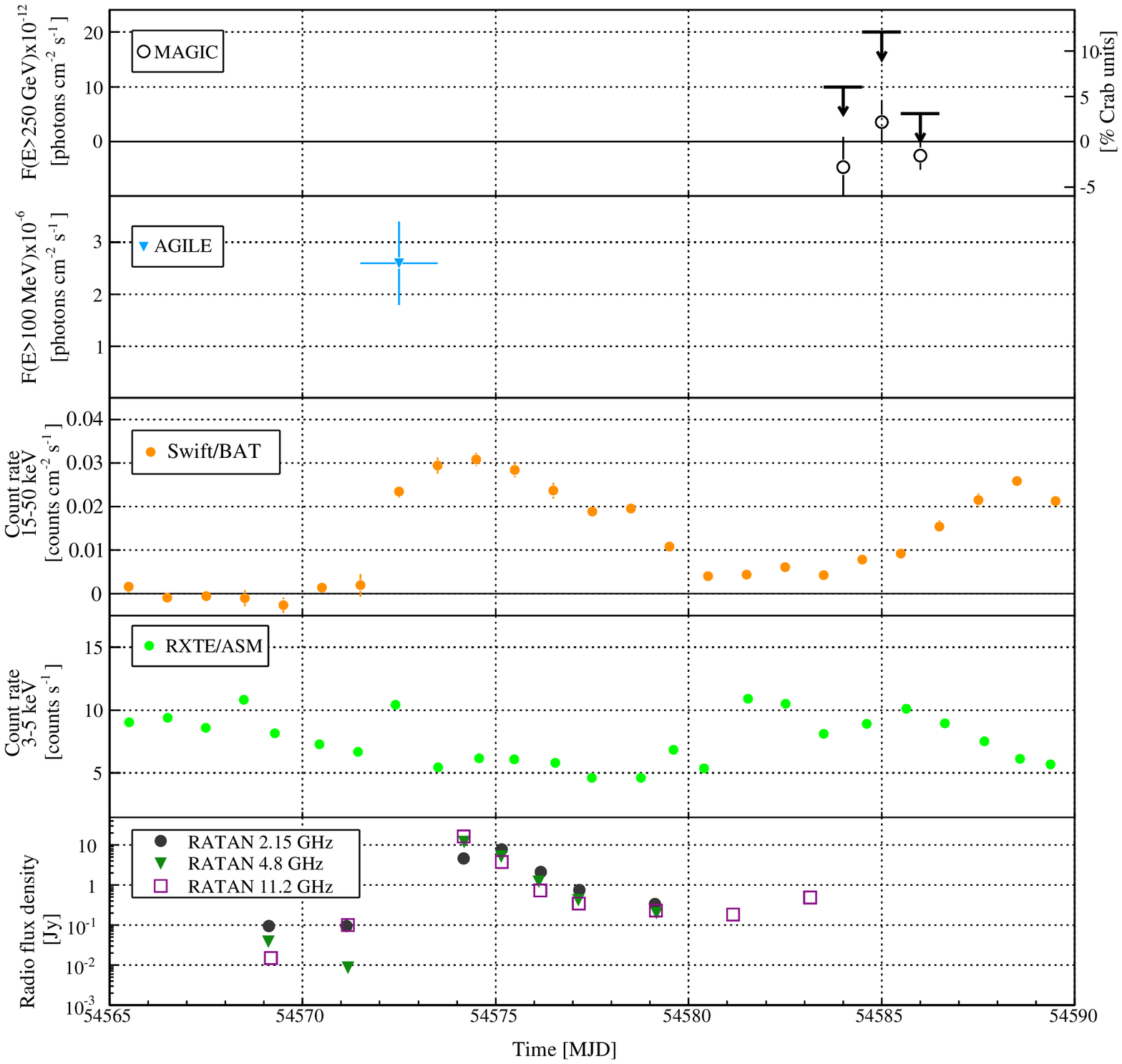} 
\figcaption{Zoom of Figure~\ref{fig::multi} around the cycle IV campaign between
2008 April 9 and May 2. The open black points in 
the VHE MAGIC panel show the non-significant measured integral fluxes with 
their statistical error bars (whereas the ULs take into account also the 
systematic errors). \label{fig::zoomIV}}
\end{center}
\end{figure*}

Figure~\ref{fig::saxo} shows the soft X-ray (3--5 keV) \emph{RXTE}/ASM 
count rates versus radio flux densities corresponding  
either to the nights of MAGIC observations for the five observational cycles 
or to the \emph{AGILE} flux peaks (only the last four \emph{AGILE} detections
were considered since no simultaneous radio data were available for the first one). 
Unfortunately, no radio data simultaneous with cycle III and cycle IV MAGIC 
observations were available. Nevertheless, in the case of cycle III data, this does 
not prevent us from pointing out that \cyg\ was in a quiescent state, just by using 
the soft X-ray measurements. For cycle IV data, 
quasi simultaneous radio observations (one day before MAGIC observations)
were used. This latter choice does not affect our qualitative result since the 
source had already entered a post-flaring state. As shown in  
Figure~\ref{fig::zoomIV}, in 2008 April, \emph{AGILE} detected \cyg\ (MJD 
54572--54573) one day before a major radio flare~\citep{Trushkin} lasting a
few days, but MAGIC started pointing at the microquasar 10 days after the 
peak radio emission.

\begin{figure*}[h]
\begin{center}
\epsscale{1.2}
\plotone {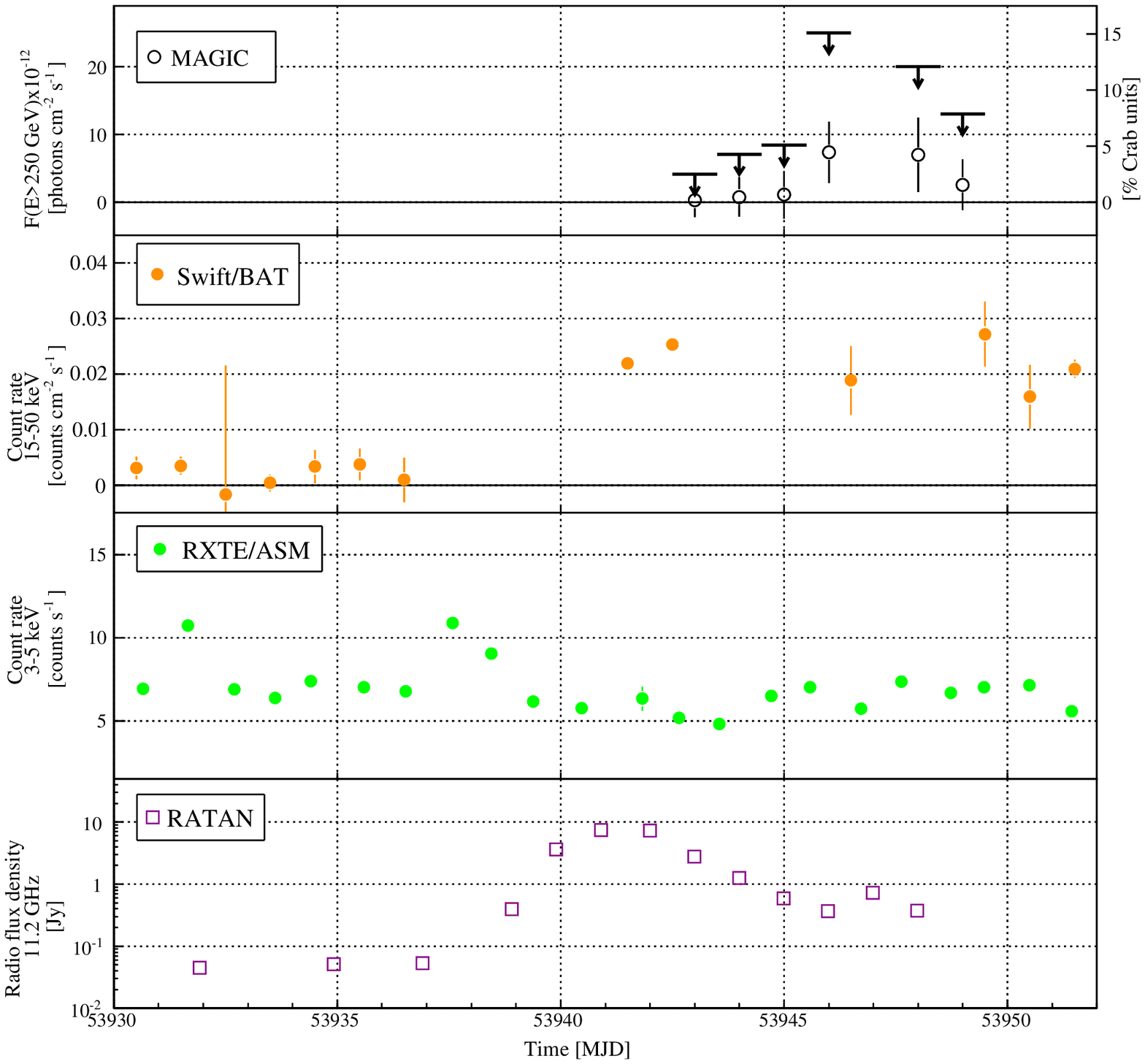} 
\figcaption{Zoom of Figure~\ref{fig::multi} around the cycle II campaign 
between 2006 July 14 and August 5. 
The open black points in the VHE MAGIC panel show the 
non-significant measured integral fluxes with their statistical error bars 
(whereas the ULs take into account also the systematic errors).
\label{fig::zoom2}}
\end{center}
\end{figure*}

All the high-energy flux peaks were detected in the right branch of the ``saxophone", 
before, after, or during a flaring state. Although \citet{Abdo2009} quoted a time 
lag between the radio and the $\gamma$-ray peaks of 5$\pm$7 days, the correlation 
between the two energy bands is not yet clear. 
On the other hand, MAGIC observed \cyg\ in its SS some days after 
the radio flare occurred (see Figures~\ref{fig::zoomIV}, \ref{fig::zoom2}), 
although for the first nights of cycle~V observations, 
the radio flux densities are rather high and oscillating between two small flares 
(see Figure~\ref{fig::zoom5}).

\section{Discussion}\label{sec::discussion}

MAGIC observations of \cyg\ cover all X-ray spectral states of the source 
in which VHE emission is thought to be likely produced either from 
a persistent jet in the HS or during powerful ejections in the SS. 
However, no significant excess events were found in any of the 
inspected samples. 

VHE $\gamma$ rays have been predicted from 
microquasar jets~\citep[]{Atoyan1999,Romero2003,Bosch-Ramon2006}.
A robust prediction of modeling is that photon--photon absorption cannot 
be neglected if the $\gamma$ rays are produced close to a massive 
star~\citep[]{Bednarek1997,Orellana2007}. In particular for \cyg\, the 
presence of a Wolf--Rayet companion, with temperature $T_*\approx 10^5$~K 
and radius $R_*\approx 10^{11}$~cm, leads to an optical depth
$\tau\ge 1$ for VHE $\gamma$ rays for an emitter located at several 
orbital radii from the star~\citep{Bednarek2010}. Even under
very efficient electromagnetic cascading, i.e., a radiation to magnetic 
energy density ratio $8\,\pi\,u_*/B^2\gg 1$, the expected VHE fluxes are 
below the sensitivity of the present instruments~\citep{Bednarek2010}.
Therefore, in order to detect VHE photons, the emitter should 
be located far from the binary system. 

\emph{Fermi}/LAT detected \cyg\ when it was in the SS and found 
orbital modulation for the radiation above $100$~MeV with a photon index of 2.7
for the periods of enhanced activity~\citep{Abdo2009}. For the epochs 
outside these high-activity periods, the GeV flux decreases 
significantly (Figure~\ref{fig::multi} in~\citet{Abdo2009}) and no modulation is found. 
The GeV orbital light curve of \cyg\ in the high-energy active periods can be 
explained in the context of anisotropic inverse Compton scattering with the 
stellar photons~\citep[]{Abdo2009,Bosch-Ramon2009,Dubus2010}, which is also 
energetically more efficient than hadronic mechanisms such as $pp$ 
interactions or photomeson production. Since very bright X-ray emission 
is produced in the inner accretion disk or at the base of the jets in 
Cygnus~X-3, the GeV radiation would be absorbed unless it is originated 
beyond $\sim 10^{10}$~cm above the compact object. This implies that the
GeV radiation is produced in the jet of \cyg\ rather than in the inner accretion 
disk/corona region. On the other hand, the GeV emitter cannot be too high in 
the jet, since otherwise there would not be strong orbital modulation 
(see also Abdo et al. 2009). Therefore,
the observed GeV and the predicted detectable VHE emission cannot be explained 
by one particular population because the former should be produced 
in/close to the system and the latter farther from it. The location of a hypothetical 
VHE emitter could coincide with the innermost region of the radio emitting jet, which, 
to avoid synchrotron self-absorption, should start relatively far from the binary system. 
This is consistent with the fact that MAGIC did not detect \cyg\ during the high-activity GeV
period, even though the flux ULs are close to a power-law 
extrapolation of the \emph{Fermi}/LAT spectrum to energies greater than 100 GeV and
well below an extrapolation of the \emph{AGILE} spectrum.

During the periods of GeV high-activity of the source, as by the synchrotron 
self Compton scenario~\citep{Atoyan1999}, a detectable 
TeV signal could arise during the first hours of a radio outburst. Unfortunately, 
MAGIC has never observed the source during this phase of the flare, but always some
days before or after the maximum radio flux. This radiation would not be 
strongly modulated, due to its origin far from the system. 
The two times more sensitive two telescopes 
arrangement, MAGIC phase II, may indeed detect
\cyg\ if it observes the source for longer time at the very 
maximum of a GeV flare, which might be even earlier than the 
onset of the radio outburst~\citep{Abdo2009}.

In the HS, the VHE emission is expected to be 
produced inside the compact and persistent jets, whose total 
luminosity is estimated to be at least 
10$^{37}$~erg~s$^{-1}$~\citep{Marti2005}. The MAGIC
VHE $\gamma$-ray UL set at 
1.1~$\times$~10$^{-12}$~erg~cm$^{-2}$~s$^{-1}$ 
is equivalent to a VHE luminosity of 
$\simeq$ 7~$\times$~10$^{33}$~erg~s$^{-1}$ at 7 kpc. 
Thus, the maximum conversion efficiency of the jet power into VHE 
$\gamma$ rays is 0.07$\%$ which is similar to that of Cygnus~X-1 
for the UL on the VHE steady emission, but 1 order of 
magnitude greater than that of GRS~1915+105~\citep[]{Albert2007b,
Acero2009}. These ULs are in good 
agreement with the theoretical expectations which generally predict 
a VHE steady luminosity of $\simeq$ 10$^{32}$~erg~s$^{-1}$.
Persistent galactic jets do not seem to be good candidate sources 
to be detected at VHEs by the current sensitivity instrumentation.
Only 10 times more sensitive future instruments, such as Cherenkov 
Telescope Array, may have a chance to detect such VHE 
emission. This would provide a new handle on the emission 
mechanisms of compact jets. 

\section{Acknowledgments}
We thank the Instituto de Astrofisica de 
Canarias for the excellent working conditions at the 
Observatorio del Roque de los Muchachos in La Palma. 
The support of the German BMBF and MPG, the Italian INFN,
the Swiss National Fund SNF, and the Spanish MICINN is 
gratefully acknowledged. This work was also supported 
by the Polish MNiSzW Grant N N203 390834, by the YIP 
of the Helmholtz Gemeinschaft, and by grant DO02-353
of the Bulgarian National Science Fund. Sergei Trushkin 
is grateful for the support of the RATAN observations
and the Russian Foundation for Basic Research,  Grant 
N08-02-00504-a.\\
We credit the \emph{Swift}/BAT and 
\emph{RXTE}/ASM teams for making public their transient 
monitor and quick-look results, respectively. 
We also thank St{\'e}phane Corbel for providing useful 
comments and information on the OVRO and 
{\it Fermi}/LAT light curves.\\



\begin{deluxetable*}{cccccccc}[!h]
\tablecaption{Differential Flux ULs for the VHE Time-integrated Emission at 95$\%$ CL.
\label{tab::DiffULtot}}
\tablehead{
\colhead{Energy Range}  & \colhead{$\mathrm{N_{on}}$ Evts.\tablenotemark{a}} & \colhead{$\mathrm{N_{bg}}$ Evts.\tablenotemark{b}} & \colhead{Excess Evts.} & \colhead{Norm.Fact.\tablenotemark{c}} & \colhead{Signif.\tablenotemark{d}} & \colhead{UL Evts.\tablenotemark{e}} &  \colhead{Flux UL}\\  
 \colhead{(GeV)}&    & & & & \colhead{($\sigma$)} & & \colhead{($\mathrm{TeV^{-1}cm^{-2}s^{-1}}$)}   
}
\startdata
199--315 & 4416 & 4384.5 $\pm$ 39.0& 31.5 $\pm$ 77.0& 0.34 & 0.4 & 237.8  & 2.6E$-$11 \\
315--500 & 2057 & 1980.6 $\pm$ 28.6& 76.4 $\pm$ 53.6& 0.36 & 1.5 & 264.2  & 1.2E$-$11\\ 
500--792 & 769 & 800.8 $\pm$ 21.1& $-$31.8 $\pm$ 34.9& 0.39 & $-$0.9 & 51.3  & 1.1E$-$12 \\
792--125 & 289 & 299.9 $\pm$ 12.9& $-$10.9 $\pm$ 21.4& 0.38 & $-$0.3 & 39.2  & 5.1E$-$13 \\
1256--1991 & 102 & 98.3 $\pm$ 6.9& 3.7 $\pm$ 12.2& 0.37 & 0.5 & 36.2 & 3.0E$-$13\\ 
1991--3155 & 38 & 32.3 $\pm$ 3.5 & 5.7 $\pm$ 7.1& 0.35 & 0.7 & 27.4 & 1.3E$-$13 
\enddata
\tablenotetext{a}{Number of signal events.\\}
\tablenotetext{b}{Number of normalized background events.\\}
\tablenotetext{c}{Normalization factor.\\}
\tablenotetext{d}{Significance.\\}
\tablenotetext{e}{Maximum number of excess events computed by using Rolke's method.}
\end{deluxetable*}

\begin{deluxetable*}{cccccccccc}[h]
\tablecaption{Integral Flux ULs for Energies Above 250 GeV Calculated
on a Daily Basis at 95\% CL.  \label{tab::ULdaily}}
\tablehead{
\colhead{Date} & \colhead{Time} & \colhead{$N_{on}$ Evts.} & \colhead{$N_{bg}$ Evts.} & \colhead{Excess Evts.}& \colhead{Norm.Fact.}& \colhead{Signif.} & \colhead{UL Evts.}&  \multicolumn{2}{c}{Flux UL} \\  
\colhead{(MJD)} &  \colhead{(h)}&  & &  &  & \colhead{($\sigma$)} & & \colhead{($\mathrm{cm^{-2}s^{-1}}$)} &\colhead{$(\%$ C.U.$)$}  
}
\startdata
53817& 0.46 & 42 & 41.4 $\pm$ 1.8 & 0.6 $\pm$ 6.7& 0.07 & 0.1 & 18.7 & 1.2E$-$11 & 7.5\\
53818& 0.26 & 43 & 27.0 $\pm$ 0.7& 16.0 $\pm$ 6.6& 0.05 & 2.8 & 46.7 & 5.7E$-$11 & 34.2\\
53820& 0.54 & 85  & 83.8 $\pm$ 4.6 & 1.2 $\pm$ 10.2 & 0.15 & 0.1 & 27.4 & 1.5E$-$11 & 9.3\\
53822& 0.70 & 132  & 111.4 $\pm$ 3.5 & 20.6 $\pm$ 12.0& 0.19 & 1.7 & 67.9 & 2.9E$-$11 & 17.4\\
53824& 0.80 & 84 & 84.8 $\pm$ 2.4& $-$0.8 $\pm$ 9.5& 0.15 & $-$0.1 & 24.3 & 9.1E$-$12 & 5.5\\
53825 & 0.80 & 83 & 64.3 $\pm$ 2.1& 18.7 $\pm$ 9.3& 0.15 & 2.1 & 58.0 & 2.2E$-$11 & 13.1\\
53826& 1.00 & 37 & 34.3 $\pm$ 0.6& 2.7 $\pm$ 6.1& 0.06 & 0.4 & 20.9 & 6.1E$-$12 & 3.7\\
53827& 0.92 & 79 & 76.5 $\pm$ 1.9& 2.5 $\pm$ 9.1& 0.13 & 0.3 & 28.4 & 9.1E$-$12 & 5.5\\
53828& 1.05 & 42 & 49.3 $\pm$ 1.1 & $-$7.3 $\pm$ 6.6& 0.09 & $-$1.0 & 10.9 & 3.1E$-$12 & 1.8\\
53943& 3.10 & 257 & 274.3 $\pm$ 9.5& $-$17.3$\pm$ 18.6& 0.33 & $-$0.9 & 27.9 & 4.1E$-$12 & 2.5\\
53944& 2.54 & 232 & 235.7 $\pm$ 8.8& $-$3.7 $\pm$ 17.6& 0.33 & $-$0.2 & 38.6 & 7.1E$-$12 & 4.3\\
53945& 1.73 & 122 & 125.7 $\pm$ 6.4& $-$3.7 $\pm$ 12.7& 0.33 & $-$0.3 & 27.1 & 8.4E$-$12 & 5.1\\
53946& 1.37 & 127 & 108.0 $\pm$ 5.9& 19.0$\pm$ 12.7& 0.33 & 1.5 & 65.0 & 2.5E$-$11 & 15.3\\
53948& 0.90 & 82 & 65.7 $\pm$ 4.6& 16.3 $\pm$ 10.2& 0.33 & 1.7 & 54.3 & 3.2E$-$11& 19.4\\
53949& 1.79 & 138 & 143.3 $\pm$ 6.9& 5.3 $\pm$ 13.6& 0.33 & 0.4 & 42.1 & 1.3E$-$11 & 7.8 \\
 54286 & 2.16 &299 &314.0 $\pm$ 10.2& $-$15.0 $\pm$ 20.1& 0.33 & $-$0.74 &62.1 &1.1E$-$11 & 6.8\\
 54287 & 4.53 &547 & 547.0 $\pm$ 13.4& $-$0.7 $\pm$ 27.0& 0.33& $-$0.03 &150.3 & 1.3E$-$11 & 8.1\\
 54288 & 1.07  &176 & 208.0 $\pm$ 8.3& $-$32.0 $\pm$ 15.6& 0.33 & $-$1.99 &20.3 & 7.0E$-$12 & 4.3\\
 54289 & 0.38 & 86 & 84.3 $\pm$ 5.3& 1.7 $\pm$ 10.7& 0.33 & 0.16 & 41.3 & 3.9E$-$11 & 23.5\\
 54295 & 1.82 & 189 & 176.3 $\pm$ 7.6& 12.7 $\pm$ 15.7& 0.33 & 0.82 &76.4 & 2.1E$-$11 & 13.0\\
 54296 & 1.93 & 219 & 209.0 $\pm$ 8.3& 10.0 $\pm$ 17.0& 0.33 & 0.59 & 74.9& 2.0E$-$11 & 11.8\\
 54297 & 1.93 & 151 & 138.7 $\pm$ 6.7& 12.3 $\pm$ 14.0& 0.33 & 0.89 &65.9 & 1.7E$-$11 & 10.4\\
 54305 & 1.75& 179 & 167.7 $\pm$ 7.4& 11.3 $\pm$ 15.3& 0.33 & 0.75 & 111.6& 3.1E$-$11 & 19.0\\
 54307 & 0.98& 115 & 117.0 $\pm$ 6.2& $-$2.0 $\pm$ 12.4& 0.33 & $-$0.16 & 52.9& 2.4E$-$11 & 14.5\\
 54308 &  0.5 & 71 & 59.3 $\pm$ 4.4& 11.7 $\pm$ 9.5& 0.33 & 1.27 & 100.1& 8.1E$-$11 & 48.8\\
 54316 & 0.73 & 85 & 88.7 $\pm$ 5.4& $-$3.7 $\pm$ 10.7& 0.33 & $-$0.34 & 33.0&1.8E$-$11 & 11.0\\
 54318 & 2.72 & 262 &277.3 $\pm$ 9.6& $-$15.3 $\pm$ 18.8& 0.33 & $-$0.81 & 39.2& 6.7E$-$12 & 4.1\\
 54319 & 1.93& 184 & 184.0 $\pm$ 7.8& 0.0 $\pm$ 15.6& 0.33 & 0.00 & 61.8& 1.5E$-$11 & 8.8\\
 54320 & 1.58 &146 & 152.3 $\pm$ 7.1 & $-$6.3 $\pm$ 14.0& 0.33 & $-$0.95 &38.8 & 1.1E$-$11 & 6.8\\
 54346 & 1.86& 182 & 200.0 $\pm$ 8.1& $-$18.0 $\pm$ 15.7& 0.33 & $-$1.12 &45.3 &1.0E$-$11 & 6.3\\
 54584 &  0.31 &17 & 23.0 $\pm$ 2.7& $-$5.0 $\pm$ 5.0& 0.33 & $-$1.0 & 8.0 & 1.0E$-$11 & 6.2\\
 54585 &  1.07 & 89 & 75.3 $\pm$ 5.0 & 13.7 $\pm$ 10.7& 0.33 & 1.3 & 50.2 & 2.0E$-$11 & 11.9\\
 54586 &  1.33 &60 & 66.7 $\pm$ 4.7 &$-$6.7 $\pm$ 9.1& 0.33 & $-$0.7 & 15.5 & 5.1E$-$11 & 3.1\\
55031& 3.48 & 186 & 183.3 $\pm$ 7.8& 2.7 $\pm$ 15.7& 0.33 & 0.2 & 43.2 & 5.7E$-$12 & 3.4\\
55033 & 2.21 & 71 & 63.7 $\pm$ 4.6 & 7.3 $\pm$ 9.6& 0.33 & 0.8 & 36.2 & 7.2E$-$12 & 4.4 \\
55034& 1.63 & 50 & 43.3 $\pm$ 3.8& 6.7 $\pm$ 8.0& 0.33 & 0.9 & 31.4 & 9.7E$-$12 & 5.8\\
55044& 1.81 & 88 & 80.0 $\pm$ 5.1& 8.0 $\pm$ 10.7& 0.33 & 0.8 & 40.0 & 9.4E$-$12 & 5.7 \\
55045& 0.88 & 69 & 69.0 $\pm$ 4.8& 0.0 $\pm$ 9.6& 0.33 & 0.0 & 24.2 & 1.1E$-$11 & 6.6
\enddata
\tablecomments{Refer to Table~\ref{tab::DiffULtot} for the meaning of the columns.}
\end{deluxetable*}

\begin{deluxetable}{ccccccccc}[t]
\tablecaption{Differential Flux ULs for the SS and HS Observations. \label{tab::DiffULspectral}}
\tablehead{
\colhead{Spectral} & \colhead{Energy Range} & \colhead{$N_{on}$ Evts.}& \colhead{$N_{bg}$ Evts.}& \colhead{Excess Evts.} & \colhead{Norm.Fact.} & \colhead{Signif.} & \colhead{UL Evts.}&  \colhead{Flux UL}\\  
\colhead{State} & \colhead{(GeV)}&    & & &  & \colhead{($\sigma$)} & & \colhead{($\mathrm{TeV^{-1}cm^{-2}s^{-1}}$)}   
}
\startdata
& 199--315 & 1709 & 1677.54 $\pm$ 24.6& 31.5 $\pm$ 48.1& 0.36 & 0.7 & 169.2  & 3.5E$-$11\\ 
& 315--500 & 926 & 858.3 $\pm$ 18.5& 67.7$\pm$ 35.6& 0.40 & 1.9 & 212.3 &  1.7E$-$11 \\
HS& 500--792 & 324 & 357.5 $\pm$ 12.9& $- $33.5 $\pm$ 22.2& 0.47 & $-$1.1 & 30.3 &  1.2E$-$12 \\
& 792--1256 & 125 & 124.9 $\pm$ 7.7& 0.1 $\pm$ 13.6& 0.48 & 0.4 & 38.0 &  8.4E$-$13 \\
& 1256--1991 & 41 & 33.7 $\pm$ 3.9	& 7.3 $\pm$ 7.5 & 0.46 & 1.4 & 32.7 &  4.5E$-$13 \\
& 1991--3155 & 14 & 8.7 $\pm$ 1.8& 5.3 $\pm$ 4.2 & 0.40 & 1.3 & 20.3 &  1.7E$-$13 \\
\tableline
 & 199--315 & 2707 & 2707.0 $\pm$ 29.9& 0.0 $\pm$ 60.0& 0.33 & 0.0 & 146.2  & 3.4E$-$11 \\
& 315--500 & 1131 & 1122.3 $\pm$ 19.2& 8.7 $\pm$ 38.7 & 0.33 & 0.2 & 108.5  & 1.1E$-$11 \\
LH& 500--792 & 445 & 443.3 $\pm$ 12.1& 1.7 $\pm$ 24.3& 0.33 & 0.1 & 62.5  & 3.3E$-$12 \\
& 792--1256 & 164 & 175.0 $\pm$ 7.5& $-$11.0 $\pm$ 14.9& 0.33 & $-$0.7 & 24.7  & 7.7E$-$13\\ 
& 1256--1991 & 61 & 64.7 $\pm$ 4.6& $-$3.7$\pm$ 9.1& 0.33 & $-$0.4 & 18.4  & 3.7E$-$13 \\
& 1991--3155 & 24 & 23.7 $\pm$ 2.7& 0.3 $\pm$ 5.6& 0.33 & 0.1 & 15.2  & 1.7E$-$13 
\enddata
\tablecomments{Refer to Table~\ref{tab::DiffULtot} for the meaning of the columns.}
\end{deluxetable}

\begin{deluxetable}{ccccccccccc}[t]
\tablecaption{Integral Flux ULs for Energies Above 250 GeV for the 
Phase-folded Analyses of the Observations in the SS and the HS.
\label{tab::PhaseULspectral}}
\tablehead{
\colhead{Spectral} & \colhead{Phase} & \colhead{Time} & \colhead{$N_{on}$ Evts.} & \colhead{$N_{bg}$ Evts.} & \colhead{Excess Evts.} & \colhead{Norm.Fact.} & \colhead{Signif.} & \colhead{UL Evts.} &  \multicolumn{2}{c}{Flux UL} \\  
 \colhead{State} & &  \colhead{(h)}&  & & &  & \colhead{($\sigma$)} & & \colhead{($\mathrm{cm^{-2}s^{-1}}$)} &\colhead{($\%$ C.U.)}  
}
\startdata
 & 0.0--0.1 & 1.34 & 64 & 68.0 $\pm$ 4.7 & $-$4.3 $\pm$ 9.3 & 0.33 & $-$0.5 & 18.2 & 6.7E$-$12 & 4.0\\
& 0.1--0.2 &1.40 & 75 & 67.3 $\pm$ 4.7 & 8.3 $\pm$ 9.8 & 0.33 & 0.9 & 38.6 & 1.5E$-$11 & 8.9\\
& 0.2--0.3& 1.63 & 108 & 107.0 $\pm$ 5.9 & 1.3 $\pm$ 12.0 & 0.33 & 0.1 & 32.1 & 9.2E$-$12 & 5.6\\
& 0.3--0.4& 2.05 & 162 & 162.0 $\pm$ 7.3 & 0.3 $\pm$ 14.6 & 0.33 & 0.0 & 37.1 & 7.7E$-$12 & 4.7 \\
& 0.4--0.5& 2.11 & 113 & 106.0 $\pm$ 5.9 & 7.3 $\pm$ 12.1 & 0.33 & 0.6 & 42.3 & 9.6E$-$12 & 5.8\\
SS & 0.5--0.6& 3.40 & 231 & 222.3 $\pm$ 8.5 & 8.7 $\pm$ 17.5 & 0.33 & 0.5 & 57.1 & 8.5E$-$12 & 5.1\\
& 0.6--0.7& 4.93 & 370 & 361.3$\pm$ 10.2 & 9.1 $\pm$ 21.8 & 0.29 & 0.4 & 68.3  & 5.7E$-$12 & 3.4\\
& 0.7--0.8& 7.13 & 649 & 600.7 $\pm$ 17.1 & 49.4 $\pm$ 30.7 & 0.49 & 1.6 & 163.9  & 8.1E$-$12 & 4.9\\
& 0.8--0.9& 4.69 & 331 & 310.0 $\pm$ 8.4 & 21.2 $\pm$ 20.0 & 0.23 & 1.1 & 86.4 & 8.3E$-$12 & 5.1\\
& 0.9--1.0& 2.05 & 109 & 116.0 $\pm$ 6.2 & $-$7.0 $\pm$ 12.1 & 0.33 & $-$0.6 & 22.1  & 5.5E$-$12 & 3.3 \\
& 0--1& 30.78 & 2216 & 2120.3 $\pm$ 29.5 & 96.3 $\pm$ 55.5 & 0.41 & 1.8 & 311.6 & 4.1E$-$12 & 2.5\\
\tableline
 & 0.0--0.1 &2.89& 511 & 509.7 $\pm$ 12.9 & 1.3 $\pm$ 26.1 & 0.33 & 0.0 &102.5 & 8.6E$-$12 & 5.2\\ 
& 0.1--0.2 & 3.23 & 477 & 501.3 $\pm$ 12.8 & $-$24.3 $\pm$ 25.3&0.33& $-$0.9 &52.5 & 4.6E$-$12 & 2.7\\
& 0.2--0.3& 2.53 & 281 & 308.0 $\pm$ 10.1 & $-$27.0 $\pm$ 19.5&0.33&$-$1.4&33.6& 4.1E$-$12&2.5\\
& 0.3--0.4& 2.53 & 266 & 230.0 $\pm$ 8.7 & 36.0 $\pm$ 18.5&0.33&2.0&218.1 & 2.1E$-$11&12.4\\
& 0.4--0.5& 2.79 & 248 & 261.0 $\pm$ 9.2 & $-$13.0 $\pm$ 18.3 &0.33&$-$0.7&43.2 & 5.4E$-$12 & 3.3\\
HS& 0.5--0.6& 2.27 & 210 & 205.3 $\pm$ 8.2 & 4.7 $\pm$ 16.6 & 0.33&0.3&73.1 & 1.1E$-$11 & 6.8\\
& 0.6--0.7& 2.99 & 242 & 243.0 $\pm$ 8.9 & $-$1.0 $\pm$ 17.9 & 0.33 & $-$0.1 & 65.8& 7.1E$-$12 & 4.3\\
& 0.7--0.8& 2.63 & 234 & 222.7 $\pm$ 8.6 & 11.3 $\pm$ 17.5 & 0.33 & 0.6 & 116.3& 1.1E$-$11 & 6.8\\
& 0.8--0.9& 1.95 & 199 & 189.7 $\pm$ 7.9 &9.3 $\pm$ 16.2 & 0.33 & 0.6 & 118.5& 1.3E$-$11 & 7.8\\
& 0.9--1.0 & 2.13 & 235 & 261.3 $\pm$ 9.3& $-$26.3 $\pm$ 17.9 & 0.33 & $-$1.4 &33.8 & 3.9E$-$12 & 2.4\\
 & 0--1& 25.9 & 2903 & 2932.0 $\pm$  31.1& $-$29.0 $\pm$ 62.2 & 0.33 & $-$0.5 & 191.0 & 1.8E$-$12 & 1.1
\enddata
\tablecomments{Refer to Table~\ref{tab::DiffULtot} for the meaning of the columns.}
\end{deluxetable}

\end{document}